# A new generalized Kohn-Sham method for fundamental band-gaps in solids

**Helen R. Eisenberg and Roi Baer\***

*Institute of Chemistry and the Fritz Haber Center for Molecular Dynamics, the Hebrew University of Jerusalem, Jerusalem 91904 Israel.*

Draft: Friday, February 06, 2009

We developed a method for calculating solid-state ground-state properties and fundamental band-gaps using a generalized Kohn-Sham approach combining a local density approximation (LDA) functional with a long-range explicit exchange orbital functional. We found that when the range parameter is selected according to the formula $\gamma = A/(\varepsilon_\infty - \tilde{\varepsilon})$ where $\varepsilon_\infty$ is the optical dielectric constant of the solid and $\tilde{\varepsilon} = 0.84$ and $A = 0.216 a_0^{-1}$, predictions of the fundamental band-gap close to the experimental values are obtained for a variety of solids of different types. For most solids the range parameter $\gamma$ is small (i.e. explicit exchange is needed only at long distances) so the predicted values for lattice constants and bulk modulii are similar to those based on conventional LDA calculations.

## I. INTRODUCTION

Density functional theory (DFT),[1] applied via the Kohn-Sham (KS) approach[2], is routinely used for the successful determination and interpretation of the structural and cohesive properties of a broad variety of solid state systems. However, solid-state band-gaps are typically under-estimated by a factor of 2. In the KS approach the system of electrons is mapped onto a system of non-interacting fermions (the KS system) governed by a local potential (in practical calculations this local potential is only approximately known). In this non-interacting system a one-particle Schrödinger equation is set up and its eigenstates and eigenvalues (called KS orbital energies) are determined. The lowest orbital energies are then used as estimates for ionization potentials. However, such a procedure is only known to be rigorously correct for the Fermi level[3,4]. The fundamental band-gap of the solid $E_g$, is often approximated by the Kohn-Sham band-gap (the difference between the bottom of the conduction band and the top of the valence band). However, this too is unjustified. For a system of $N$ electrons the fundamental band-gap is a ground state quantity in the sense that it can be expressed in terms of ground-state energies, $E_{gs}$, of the $N-1$, $N$ and $N+1$ electron systems as follows:

$$E_g = IP - EA \\ = \lim_{N\to\infty} E_{gs}(N-1) - 2E_{gs}(N) + E_{gs}(N+1) \quad (1.1)$$

where $IP$ is the ionization potential and $EA$ the electron affinity. For a finite electron system the band-gap can be computed by considering the KS systems of N and N+1 particles and can be expressed as follows:

$$E_g = -\varepsilon_{N,N} + \varepsilon_{N+1,N+1} = (\varepsilon_{N,N+1} - \varepsilon_{N,N}) + \Delta_{XC} \quad (1.2)$$

Where $\varepsilon_{N,M}$ is the $M^{th}$ KS orbital energy of the N-particle KS system and $\Delta_{XC} = \varepsilon_{N+1,N+1} - \varepsilon_{N,N+1}$ is called the derivative discontinuity (because it does not go to zero in the $N \to \infty$ limit) and the term in parenthesis is the KS band-gap. Evidently, even if the exact local potential of the KS systems was known, the common procedure of taking the KS band-gap would not give the correct solid-state gap as one would still need to add to the orbital gap the derivative discontinuity $\Delta_{XC}$.[5-7]

The popular local and semi-local approximations, such as the local density approximation (LDA) and the generalized gradients approximations (GGAs), are believed to give local potentials and orbital energies which are not bad approximations to the exact KS quantities.[8] However because of their inherent semilocal density structure, they predict a zero derivative discontinuity ($\Delta_{XC} = 0$).[5] Therefore the fundamental band-gaps inferred from them, are similar to the exact Kohn-Sham band gaps but are poor approximations to the experimental gaps (by a factor ~2). The fact that derivative discontinuities are missing from semi-local functionals has been associated with the existence of spurious electron self-repulsion.[5,9,10]

The complication of adding the derivative discontinuity to the KS band can be circumvented in the generalized Kohn-Sham (GKS) framework[11] which uses explicit orbital functionals (i.e the orbitals are explicitly contained in the functional as opposed to the functional being only explicitly dependent on the density). With explicit orbital functionals, the orbital energy band-gap already incorporates some or all of the derivative discontinuity and so may be used to directly approximate the experimental band-gap. For example, if the orbital functional includes a Hartree-Fock like exchange operator the derivative discontinuity $\Delta_{XC}$ can be decomposed into the sum of corresponding exchange ($\Delta_X$) and correlation ($\Delta_C$) parts. It was demonstrated, using perturbative arguments, that the orbital energy gap in the GKS equation incorporates much of $\Delta_X$ while the correlation derivative discontinuity $\Delta_C$ was estimated as a much smaller correction.[11]

---

\* To whom correspondence should be sent. Email: roi.baer@huji.ac.il.

There are several previous works which use a generalized KS framework for band-gaps. One approach splits the exchange energy into an explicit short-range exchange operator and a local functional for the long-range exchange.[11-13] This resulted in significant improvements over the LDA fundamental band-gap for some of the materials studied. A similar approach was applied using the HSE functional with much improved results, probably due to the use of more advanced semi-local exchange-correlation functionals.[14] Both these approaches are not expected to include the entire exchange discontinuity $\Delta_X$ because the orbital functional they use does not include a long-range self-repulsion correction. Therefore, one cannot expect the fundamental band-gap to be fully contained in their GKS orbital gap. This problem can be fixed by using an explicit exchange operator for the long-range (instead of the short range) part of the exchange. This might thus be a more appropriate way to circumvent the derivative discontinuity contribution. Such an attempt was reported recently but huge gaps were reported for most materials.[15]

In this work we present a generalized Kohn-Sham method which deploys an explicit orbital exchange that incorporates the long-range self-repulsion correction exactly. The procedure introduces into the DFT correlation energy expression a short-range but non-local exchange-like orbital functional which eliminates the detrimental effects of the full "Hartree-Fock-like" exchange.[16-19] In this sense our approach is similar to that of Ref. [15]. The crucial difference is that here the range-parameter $\gamma$ is not considered "universal". It depends on the density of the system and must be tuned separately for each system[19-22]. We show how such a "tuning" can be done for solid state systems: the range-parameter $\gamma$ was found empirically to correlate very well with the optical dielectric constant $\varepsilon_\infty$ of the solid. Based on this relation we developed a method that describes the usual ground-state properties of solids (lattice parameter and bulk modulus) with LDA quality while simultaneously describing the fundamental band-gaps extremely well.

Our theory for the range-separated hybrid is discussed in section II. The relation between the range parameter and the dielectric constant is the topic of section III. The performance of the method is demonstrated in section IV followed by summary and discussion in section V.

## II. THE RANGE-SEPERATED HYBRID FUNCTIONAL

In the Kohn-Sham approach DFT the ground-state energy of a system of $N_e$ electrons in an external potential $v(\mathbf{r})$, with particle-density $n(\mathbf{r})$ and a many-electron wave function $\psi_{GS}$ is expressed using quantities calculated for a system of *non-interacting fermions* with an identical ground-state density. The wave function of this "non-interacting" system is a Slater determinant and is denoted by $\psi_S$.[2] Both wave functions, $\psi_{GS}$ and $\psi_S$ are unique (up to global phase) functionals of the density. The energy of the interacting system is written as:

$$E = \langle \psi_{GS} | \hat{H} | \psi_{GS} \rangle = \langle \psi_S | \hat{H} | \psi_S \rangle + E_C[n] \quad (2.1)$$

Where $\hat{H} = \hat{T} + \hat{V} + \hat{U}$ is the Hamiltonian of the interacting system and $\hat{T} = \sum_{n=1}^{N_e}\left(-\frac{1}{2}\nabla_n^2\right)$, $\hat{V} = \sum_{n=1}^{N_e} v(\mathbf{r}_n)$ and $\hat{U} = \frac{1}{2}\sum_{n\neq m}^{N_e} \frac{1}{r_{nm}}$ are respectively the kinetic energy, the interaction with the external potential and the electron Coulomb repulsion operators (in atomic units). The last term $E_C[n]$ is the negative correlation energy functional which, because the two systems have the same density, can be written as:

$$E_C[n] = \langle \psi_{GS} | \hat{T} + \hat{U} | \psi_{GS} \rangle - \langle \psi_S | \hat{T} + \hat{U} | \psi_S \rangle \quad (2.2)$$

As we explain below, there exists a $\gamma$ for which the correlation energy, can also be expressed as:[19]

$$E_C[n] = \langle \psi_{GS} | \hat{Y}_\gamma | \psi_{GS} \rangle - \langle \psi_S | \hat{Y}_\gamma | \psi_S \rangle . \quad (2.3)$$

where $\hat{Y}_\gamma$ is a shielded Coulomb interaction energy:

$$\hat{Y}_\gamma = \frac{1}{2}\sum_{n\neq m} y_\gamma(|\mathbf{r}_n - \mathbf{r}_m|) \quad (2.4)$$

With the pair potential:

$$y_\gamma(r) = \frac{erfc(\gamma r)}{r} \qquad r > 0 \quad (2.5)$$

This function has the properties that $\lim_{\gamma\to\infty} y_\gamma(r) = 0$ and $y_0(r) = \frac{1}{r}$, which can be used to show that for each ground-state density $n(\mathbf{r})$ there exists a $\gamma$ for which Eq. (2.3) holds exactly.[19]

Even when $\gamma$ is known expression (2.3) is not practical for calculations since we have no access to $\psi_{GS}$. We thus follow the spirit of the local density approximation and approximate it as:

$$E_C^{Hyb}[n] = \int f_{XC}^\gamma(n(\mathbf{r})) d^3r \\ + \frac{1}{2}\int |\rho(\mathbf{r}_1,\mathbf{r}_2)|^2 y_\gamma(|\mathbf{r}_1-\mathbf{r}_2|) d^3r_1 d^3r_2 \quad (2.6)$$

in which $\rho(\mathbf{r}_1,\mathbf{r}_2)$ is the density matrix of the non-interacting

system and

$$f_{XC}^\gamma(n(\mathbf{r}),|\nabla n(\mathbf{r})|,...) = [\varepsilon_C^{HEG}(n(\mathbf{r})) + \varepsilon_X^\gamma(n(\mathbf{r}))]n(\mathbf{r}) \quad (2.7)$$

$\varepsilon_C^{HEG}(n)$ is the correlation energy per electron of the homogeneous gas (HG) of electrons at density $n$ and $\varepsilon_X^\gamma$ is a the exchange energy of the HG of particles interacting with the potential $y_\gamma(\mathbf{r})$.[23] This approach was depicted in more detail in Ref.[19] Similar approaches deploying a system-independent $\gamma$ were conceived earlier.[16-18, 24] Recently a similar method to the one described here was used for those molecules which are problematic for conventional DFT.[20] In that paper it was shown that ab-initio tuning of the range parameter is necessary in order to describe the symmetric radical cation R+R$^+$ in the ground state.[21] Furthermore, range parameter tuning is also required for charge-transfer excitations.[25]

## III. APPLICATION FOR SOLIDS

### A. Technical computational details

The previous section has briefly reviewed the range separated functional; we now describe its use for predicting the properties of solids. The calculations described below were carried out using the Quantum-ESPRESSO package[26] which we modified to include the functional described in Eqs. (2.6)-(2.7). The local correlation functional used was the PZ81 LDA functional.[27] The local exchange functional is the LDA exchange of a homogeneous gas of particles interacting with the $y_\gamma(r)$ pair potential.[16, 18] All calculations used a plane-wave basis and were converged for kinetic energy cutoff for wave-functions (varied with material) and **k**-point grid density (4x4x4 grid was used). Calculations were carried out at the minimum energy lattice constant for each value of $\gamma$. Norm-conserving pseudopotentials based on the PZ81 exchange-correlation functional were used and all the calculations were fully self-consistent.

We ignored spin-orbit (SO) splitting effects when calculating the fundamental band-gap. These effects are negligible in all the systems we studied except AlSb and to a lesser extent AlAs. Following ref.[11] the correction to the DFT calculated fundamental gap due to spin-orbit splitting is approximately $-\Delta_{SO}/3$ at $\Gamma$ (where $\Delta_{SO}$ is the spin-orbit splitting energy). For AlSb this gives a SO correction of -0.23eV (13% of the band gap) and for AlAs a SO correction of -0.09eV (4% of the band gap). In future works we intend to investigate the effects of large SO splitting and determine whether or not corrections for this need to be included in our method.

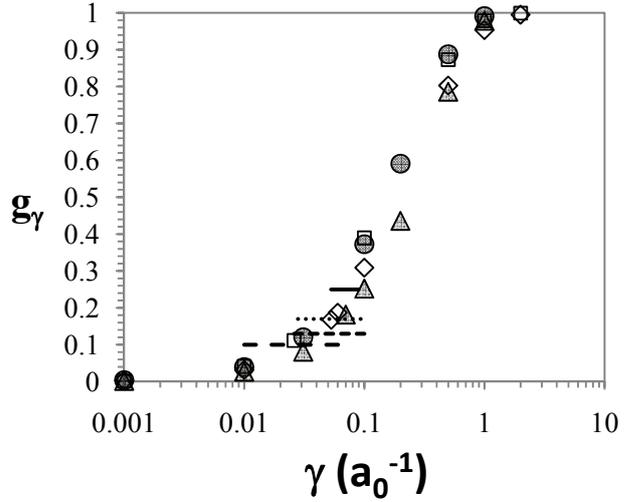

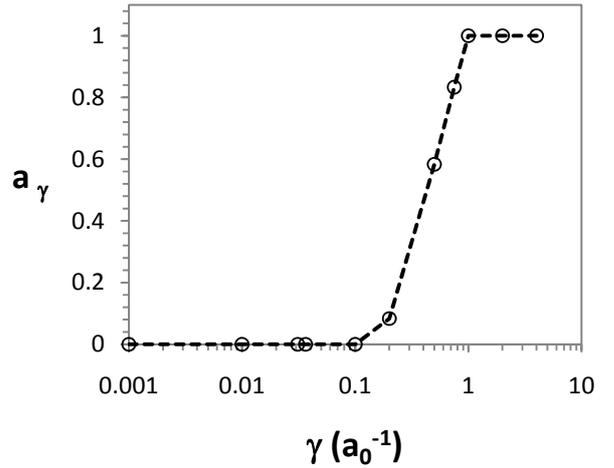

Figure 1: Top: The normalized band gap (Eq. (3.1)) as a function of $\gamma$ for several materials. Direct (indirect) band gaps are denoted "dir" ("indir"). The horizontal lines are the experimental results (exp). Bottom: the normalized lattice constant (Eq. (3.2)) of SiC.

### B. Band-gap dependence on the range parameter

As also found for molecules[20, 21, 25] the range parameter $\gamma$ in the functional must undergo a tuning stage. In order to determine a method for this we study the dependence of the fundamental-gap on $\gamma$ by considering the following non-dimensional quantity:

$$g_\gamma = \frac{E_g^\gamma - E_g^0}{E_g^\infty - E_g^0}, \quad E_g^0 \equiv E_g^{LDA}, \; E_g^\infty \approx E_g^{HF}, \quad (3.1)$$

where $E_g^\gamma$ is the gap computed with the range-separated hybrid functional with parameter $\gamma$. By definition, $g_\infty = 1$ and this was found to be close to the results of Hartree-Fock (HF) theory, while $g_0 = 0$ is the LDA limit for this equation. We plot the values of $g_\gamma$ for several materials, diamond (C), Silicon, SiC, and MgO in the top panel of Figure 1. It can be seen that the normalized gap energy is sensitive to $\gamma$ for $0.01 < \gamma < 1$. The experimental band gap values typically coincide with calculated ones in the range $0.01 < \gamma < 0.1$. A different behavior is seen for the lattice constant $a$: it is insensitive to $\gamma$ when $0.01 < \gamma < 0.1$ and takes the LDA value. This is clearly seen in the bottom panel of Figure 1, where the normalized lattice constant, defined by:

$$a_\gamma = \frac{a^\gamma - a^0}{a^\infty - a^0}, \qquad a^0 \equiv a^{LDA}, \; a^\infty \approx a^{HF}, \qquad (3.2)$$

is shown for SiC (other materials show similar behavior) as a function of $\gamma$. Change in the lattice constant starts at $\gamma > 0.1$.

We conclude that the material band-gaps are very sensitive to $\gamma$ in the experimentally relevant range while the lattice constants take the LDA values there. Furthermore each material requires a different value of $\gamma$ in order to obtain the experimental band-gap. For any given material we designate by $\gamma^*$ the value of $\gamma$ for which the calculated fundamental band-gap is equal to the corresponding experimental band-gap.

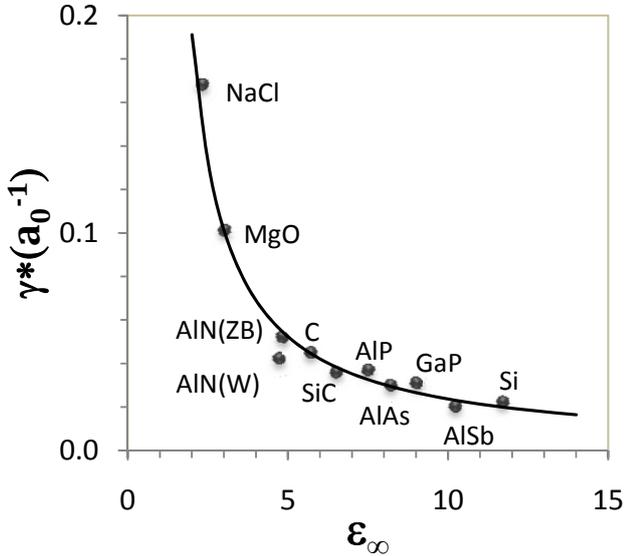

Figure 2: The value of $\gamma$ giving experimental fundamental band-gaps, $\gamma^*$, plotted against the experimental value of the optical dielectric constants $\varepsilon_\infty$ for different materials. The line depicts the relation given in Eq. (3.3) with $A = 0.216 a_0^{-1}$ and $\tilde{\varepsilon} = 0.84$.

## C. The relation between $\gamma^*$ and the dielectric constant

The basic problem is to find a method for determining $\gamma^*$. Thus, we have to find an additional relation between $\gamma$ and another internal property of the solid, which can be calculated from the electronic structure. This way one can determine $\gamma^*$ computationally without any material-dependent external input parameters.

We found that a simple relation exists between $\gamma^*$ and the optical dielectric constant $\varepsilon_\infty$. We plot in Figure 2 the value of $\gamma^*$ as a function of the experimental $\varepsilon_\infty$ for several solids. It can be seen that the relation between $\gamma^*$ and $\varepsilon_\infty$ can be approximately described by a simple functional relation with two semiempirical parameters $A$ and $\tilde{\varepsilon}$:

$$\gamma^*(\varepsilon_\infty) \approx \frac{A}{\varepsilon_\infty - \tilde{\varepsilon}} \qquad (3.3)$$

This relation will naturally give $\gamma^* \to 0$ for metallic systems (where $\varepsilon_\infty \to \infty$), therefore LDA is recovered which is appropriate for metals. In order to avoid singularities and negative values of $\gamma^*$, $\tilde{\varepsilon}$ must be smaller than 1 (all materials have $\varepsilon_\infty \geq 1$). We fitted to the data the constants $A = 0.216 a_0^{-1}$ and $\tilde{\varepsilon} = 0.84$. We adopt these values as a semipirical relation and use them in this study.

**Algorithm.** The way to use Eq. (3.3) in a calculation is:

(1) Perform an LDA calculation to determine $\varepsilon_\infty$ and the k-point values $\mathbf{k}_{val}$ of the maximum of the valence band and $\mathbf{k}_{cond}$ of the minimum of the conduction band.

(2) Determine $\gamma^*$ based on $\varepsilon_\infty$ from Eq. (3.3) and perform the DFT calculation with the range-separated hybrid functional with $\gamma^*$.

(3) Determine the band-gap from:

$$E_g = \varepsilon_{cond}(\mathbf{k}_{cond}) - \varepsilon_{val}(\mathbf{k}_{val}), \qquad (3.4)$$

where the $\varepsilon$'s are the orbital energies.

For small $\gamma$ the calculated dielectric constant $\varepsilon_\infty$ does not significantly depend on $\gamma$. We plotted the ratio $r(\gamma) = \frac{\varepsilon_\infty(\gamma)}{\varepsilon_\infty(0)}$ as a function of $\gamma$ for Si and MgO (Figure 3). It seen that for Si the computed $\varepsilon_\infty$ does not change significantly from its LDA value for $\gamma < 0.05$ and even until $\gamma = 0.1$ the change is small (less than 6%). For MgO, the range at which $\varepsilon_\infty$ takes the LDA value is even larger. We conclude that for materials with small $\gamma^*$ (such as semiconductors) the proximity of the hybrid functional to LDA justifies using the LDA value of $\varepsilon_\infty$. For larger gap materials this may need to be checked on a case by case basis and if needed the algorithm above can be modified to include an iteration step requiring self-consistency between $\varepsilon_\infty$ and $\gamma^*$.

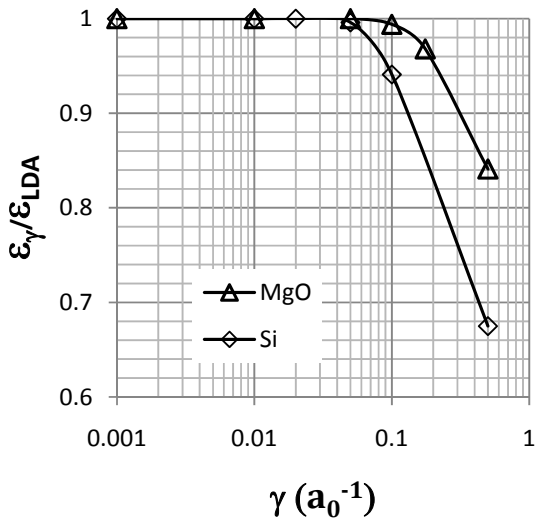

Figure 3: The dependence of the calculated dielectric constant on $\gamma$.

An important point to note is that semiconducting systems (which are close to metals) have very small $\gamma^*$ values, and so the predictions of the hybrid functional concerning ground-state properties become almost identical with those of LDA. Even the large band-gap systems, such as NaCl have a value of $\gamma^*$ smaller by more than a factor 2 than the typical values of $\gamma$ used for gas-phase molecules (which have $\gamma \sim 0.5$[20]). Concerning the wide gap materials we refer the reader to a discussion in the next subsection concerning the effects of large $\gamma$ values on the band structure.

### D. The effect of the range-parameter on the band structure

The method discussed in the previous subsection involves the use of the LDA dielectric constant $\varepsilon_\infty$ to determine $\gamma^*$ and the orbital energies of the $\gamma^*$-hybrid functional evaluated at the LDA k-points $\mathbf{k}_{val}$ and $\mathbf{k}_{cond}$ for estimating the fundamental band-gap. We found that in ionic systems the LDA fundamental band-gap $\Gamma_v \to \Gamma_c$ is closely followed by a slightly higher gap at $k_v = (0, 0, 0.5) \to k_c = (0, 0, 0)$ (where the k-points are expressed in Cartesian coordinates in units of $2\pi/a$, where $a$ is the lattice constant used in the calculation in atomic units). We show this in Figure 4 for MgS: the LDA $\Gamma_v \to \Gamma_c$ gap is 3.2 eV and the gap is 3.8 eV. The two gaps respond differently when $\gamma$ is increased. The $\Gamma_v \to \Gamma_c$ gap increases with $\gamma$ while the $(0, 0, 0.5) \to (0, 0, 0)$ gap is resistant. As a result, the hybrid functional for $\gamma > 0.02$ predicts that the $\Gamma_v \to \Gamma_c$ gap ceases to be fundamental. The problem can be traced to the distortion of the valence band when $\gamma$ changes, while the conduction band remains relatively unchanged. A similar phenomenon happens for the other ionic solids we checked (MgO and NaCl). In non-ionic solids the band structure is not as sensitive to $\gamma$ and this problem was not seen at small values of $\gamma$.

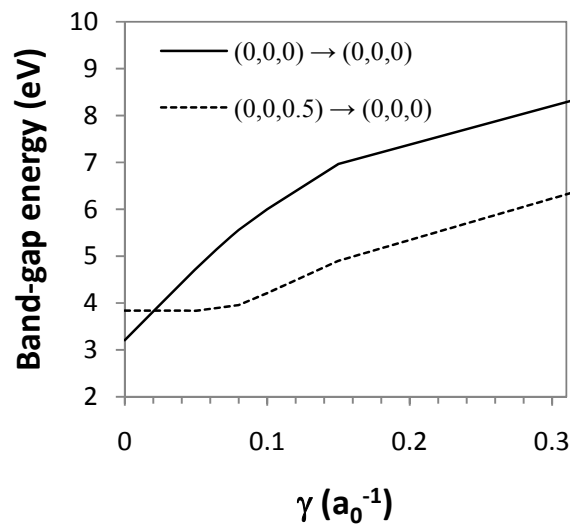

Figure 4: The change in the $(0, 0, 0.5) \to (0, 0, 0)$ and the $(0, 0, 0) \to (0, 0, 0)$ ($\Gamma_v \to \Gamma_c$) band-gaps of MgS as a function of the range parameter $\gamma$ The k-points are expressed in Cartesian coordinates in units of $2\pi/a$, where $a$ is the lattice constant used in the calculation in atomic units. The experimental band gap of MgS is 4.9 eV[28].

### IV. RESULTS

Our generalized KS DFT method consists of using as a correlation functional Eqs. (2.6)-(2.7) with the value of $\gamma^*$ determined by the relation (3.3). The dielectric constant in the latter equation is to be taken from the LDA calculation. However, we used the experimental dielectric constant instead in this paper (which are expected to be close to those of LDA) in order to demonstrate the concepts. We further discuss this issue later, and give a few results with calculated dielectric constants.

We begin with ground-state properties: the lattice constant $a$ and the isotropic bulk modulus $B$. We computed $E(a)$, the SCF energy for several lattice constants near the optimal $a_{min}$; we then fit to the resulting data a low order polynomial $E_p(E) = b_0 + b_1 E + b_2 E^2 + ...$, using the polynomial we determined the exact value of the lattice constant $a_{min}$ (where $E'_p(a_{\min}) = 0$) and for the Bulk modulus we computed the 2$^{nd}$ derivative $E''_p(a_{\min})$ from which the bulk modulus is determined by: $B = \frac{a^2 E''_p(a_{\min})}{9V}$ where $V$ is the unit cell volume. For several materials, where anharmonic effects are significant, we noticed that the second derivative results were somewhat sensitive to the order and location of data points we used; thus we determined a confidence interval for the value of B. The results of this procedure are depicted in Table 1 where the calculation results for the lattice constants and the bulk modulii are compared with various experimental measurements. The lattice constants we computed were almost identical with those of LDA. This is a result of the very small value of $\gamma^*$ which we used which leaves us very close to the LDA limit. As for the bulk modulus, this quantity is more sensitive to $\gamma^*$ however in all cases we obtained values which are in agreement (or very closely so) with experimental inaccuracies. The largest discrepancy was for NaCl where our result although too large by 20% was somewhat better than LDA.

Table 1: Dielectric constants, lattice constants and bulk modulii for the various solids considered in this study with comparison to experiment.

| Material | Symmetry | Solid type | Experimental $\epsilon_\infty$[a] | $\gamma^*(\epsilon_\infty)$ ($a_0^{-1}$) | Lattice constant ($a_0$) | | | Bulk Modulus (GPa) | | |
|---|---|---|---|---|---|---|---|---|---|---|
| | | | | | LDA | LDA-$\gamma$ | Experiment[b] | LDA | LDA-$\gamma$ | Experiment[d] |
| C | Diamond; | Insulator | 5.7 | 0.045 | 6.67 | 6.67 | 6.74 | 580-585 | 510-640 | 440-560 |
| Si | Diamond; | Semicond. | 11.7 | 0.020 | 10.23 | 10.23 | 10.26 | 96-97 | 92 | 80-100 |
| SiC | Zinc-blende | Semicond. | 6.52 | 0.038 | 8.21 | 8.22 | 8.24 | 195-245 | 215-225 | 220-260 |
| AlAs | Zinc-blende | Semicond. | 8.16 | 0.031 | 10.61 | 10.61 | 10.68 | 72-75 | 74-76 | 74 |
| AlP | Zinc-blende | Semicond. | 7.5 | 0.033 | 10.23 | 10.23 | 10.31 | 89 | 89 | 86[e] |
| AlN | Zinc-blende | Semicond. | 4.84 | 0.054 | 8.07 | 8.08 | 8.28 | 187-196 | 218 | 191-218 |
| AlN | Wurtzite | Semicond. | 4.66 | 0.057 | a 5.76 | 5.76[c] | 5.88 | 200-220 | 200-220 | 160-210 |
| | | | | | c 9.10 | 9.10[c] | 9.41 | | | |
| | | | | | u 0.388 | 0.388[c] | 0.382 | | | |
| AlSb | Zinc-blende | Semicond. | 10.24 | 0.023 | 11.55 | 11.55 | 11.58 | 56 | 56 | 55-57 |
| GaP | Zinc-blende | Semicond. | 9.0 | 0.027 | 10.10 | 10.10 | 10.28 | 93 | 93 | 85-91[f] |
| MgO | Cubic (rock salt) | Ionic solid | 2.95 | 0.104 | 7.94 | 7.95 | 7.96 | 170 | 163-165 | 160-165[g] |
| NaCl | Cubic (rock salt) | Ionic solid | 2.25 | 0.157 | 10.83 | 10.86 | 10.58 | 30-38 | 28-31 | 24[h] |

Comments:
(a) C, Si, SiC, AlAs,[29] AlP,[30] AlN,[31,32] AlSb,[33] GaP - Landolt-Börnstein III/41, MgO[34], NaCl[35]
(b) C, Si, SiC,[29] AlAs, AlP, AlN, AlSb, GaP[36], MgO[37], NaCl[38]
(c) For the wurtzite structure we took the LDA lattice parameters for u and c/a and minimized for a.
(d) Taken from Landolt-Börnstein III/41 unless where stated.
(e) Ref [39]
(f) Ref [40]
(g) Refs. [41-43]
(h) Ref [44,45]

Next we consider the fundamental band-gaps. The values of $\gamma^*(\epsilon_\infty)$ shown in Table 1 were used to compute the fundamental band-gaps of different materials. We found good band-gaps, as shown in Figure 5. This is not very surprising in view of the nice fit Eq. (3.3) gives for the data of $\gamma^*$. The only point lying significantly below the curve of Figure 2 is that corresponding to AlN in the Wurzite configuration (AlN(Wur)). Indeed the band-gap predicted for this system deviates more significantly from the experimental band-gap than in the other systems.

In all cases the band gaps, which are much too small according to LDA, were greatly improved according to the new functional. Especially encouraging is the fact that small semiconductor gaps are described as well as large ionic gaps using the simple 2-parameter formula in Eq. (3.3).

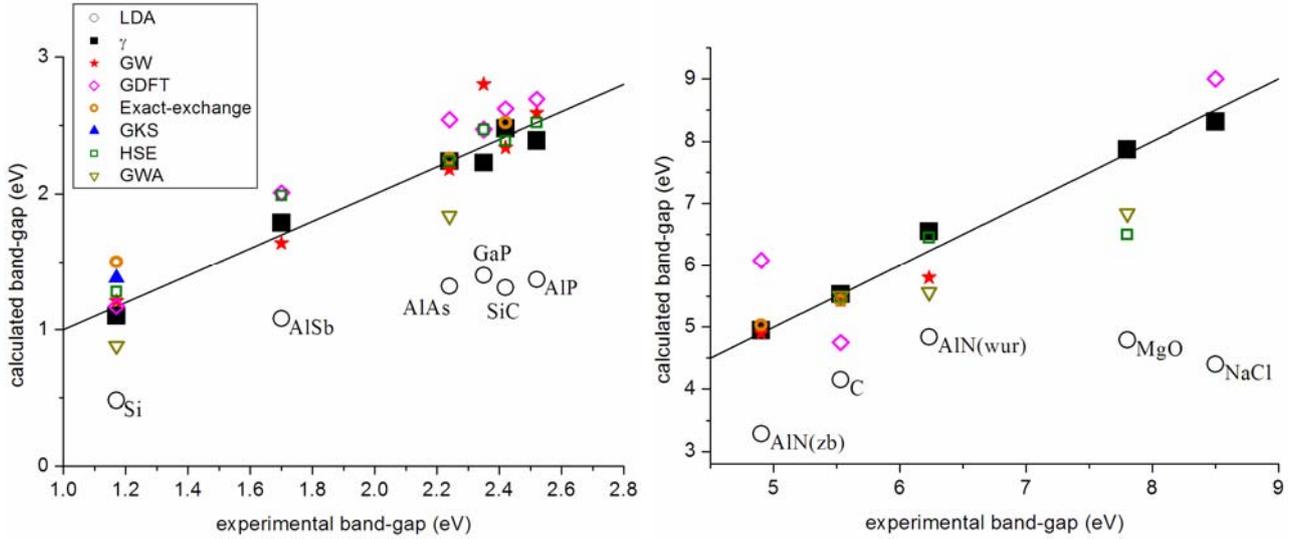

Figure 5: Comparison of calculated vs experimental fundamental band-gaps using different methods. The straight line shows the experimental results. LDA refers to the results obtained using Quantum-ESPRESSO[26] in the local density approximation; γ refers to the results obtained using our modified Quantum-ESPRESSO including our new functional; GW refers to results obtained using the GW approximation: AlN,[46] Si,[47] C,[48] SiC,[49] AlAs,[50, 51] GaP,[52] AlP, AlSb,[53]; GWA refers to results obtained from the GW approximation[54], GDFT refers to results obtained using generalized-density-functional theory: Si, C, SiC, AlAs, AlP, AlN-zinc-blende, AlSb, GaP;[55] NaCl[56], Exact-exchange refers to results obtained using the exact-exchange Kohn-Sham formalism[57]; GKS refers results obtained using the Generalized Kohn-Sham scheme[11]; HSE refers to results obtained using the Heyd-Scuseria-Ernzerhof screened hybrid potential[14].

A detailed comparison of the relative band gap errors is shown in Table 2. The LDA band gaps are ~40% too small, as is well known. GDFT and exact exchange typically over estimates the gaps by nearly 10%. HSE and GW show a good balanced description as they exhibit small mean relative errors but the mean absolute relative error is 5-6%. The new functional is also very balanced, showing vanishing mean relative errors and the mean absolute relative error is small as well, 3%.

Table 2: Statistics on relative errors in the calculated fundamental band-gaps for the various methods considered here. MeanAbs refers to the mean absolute relative error.

| Method | N | Mean | MeanAbs |
|---|---|---|---|
| γ | 10 | 0% | 3% |
| GW[46-53] | 9 | 1% | 5% |
| HSE[14] | 9 | 2% | 6% |
| Exact-exch[57] | 5 | 7% | 8% |
| GDFT[55, 56] | 8 | 8% | 11% |
| GWA[54] | 5 | -13% | 13% |
| GKS[11] | 1 | 18% | 18% |
| LDA | 10 | -39% | 39% |

LDA dielectric constants are somewhat higher than experiment. For example, in Si the experimental dielectric constant is $\epsilon_\infty^{exp} = 11.7$ while our LDA calculation yielded $\epsilon_\infty^{LDA} = 13.1$. This is a relative error of ~12%. This does not mean that our estimate of the band gap has a relative error of 12%. It is about a factor of 2 smaller as can be explained by the following considerations: Using the LDA dielectric constant in Eq. (3.3) will lead to smaller than optimal $\gamma^*$, by ~12%. This will cause the calculated normalized band gap $g_\gamma$ (Eq. (3.1)) to decrease by ~12% (see Figure 1). But the effect on the actual gap is seen to be (from Eq. (3.1)):

$$\frac{\delta E_g^\gamma}{E_g^\gamma} = \left(1 - \frac{E_g^{LDA}}{E_g^\gamma}\right)\frac{\delta g_\gamma}{g_\gamma} \qquad (4.1)$$

In Si $\left(1 - E_g^{LDA}/E_g^\gamma\right) \approx 0.6$ (i.e. LDA band gaps have a relative error of 60% with respect to the true band-gap) and so we expect the relative error in the gap, according to the new functional based on the LDA dielectric constant, to be about ~7%. In MgO, where the LDA dielectric constant is closer to experiment the same type of calculation leads to a much smaller error, to 2.5%. In a future publication we will investigate the relation between the LDA dielectric constant and the experimental gaps more closely.

## V. SUMMARY AND DISCUSSION

We developed a theory for applying the range separated hybrid developed in ref.[19] for solid-state systems. We showed that the range parameter $\gamma$ must be selected according to a simple empirical equation depending on the optical dielectric constant of the material (Eq. (3.3)). This leads to a new method which predicts ground-state properties (lattice constants, bulk modulii) with an accuracy comparable to that of LDA and at the same time gives band-gaps which are close to experimental values. We have demonstrated the results on a series of systems spanning semiconductors, insulators and wide band gap materials such as ionic solids. We showed that this method is self-contained within the modified DFT code and does not rely on any external material-dependent input parameters.

We intend to test our new method on additional classes of solids, including those with large spin-orbit splitting. In ionic solids we found that the range parameter significantly distorts the band structure.. The solution for this problem will be a primary future direction of our research and may require developing a method including a k-dependent range parameter. Such an approach will probably be necessary also for addressing, within a time-dependent DFT approach optical properties of solids beyond the fundamental gap (i.e the band-structure). Finally, an additional future research direction is the applicability to reduced dimensionality infinite systems (1D polymers and 2D sheets) and large but finite systems, e.g. proteins.

**Acknowledgements:** We gratefully acknowledge support of the US-Israel Binational Science Foundation.

[1] P. Hohenberg and W. Kohn, Phys. Rev **136**, B864 (1964).
[2] W. Kohn and L. J. Sham, Phys. Rev **140**, A1133 (1965).
[3] M. Levy and J. P. Perdew, in *Density Functional Methods in Physics*, edited by R. Dreizler and J. Perovidencia (Plenum, New York, 1985), p. 11.
[4] C.-O. Almbladh and U. von-Barth, Phys. Rev. B **31**, 3231 (1985).
[5] J. P. Perdew, R. G. Parr, M. Levy, and J. L. Balduz, Phys. Rev. Lett. **49**, 1691 (1982).
[6] L. J. Sham and M. Schluter, Phys. Rev. Lett. **51**, 1888 (1983).
[7] J. P. Perdew and M. Levy, Phys. Rev. Lett. **51**, 1884 (1983).
[8] K. Burke, J. P. Perdew, and M. Ernzerhof, J. Chem. Phys. **109**, 3760 (1998).
[9] S. Kümmel and L. Kronik, Rev. Mod. Phys. **80**, 3 (2008).
[10] A. J. Cohen, P. Mori-Sanchez, and W. T. Yang, Science **321**, 792 (2008).
[11] A. Seidl, A. Gorling, P. Vogl, J. A. Majewski, and M. Levy, Phys. Rev. B **53**, 3764 (1996).
[12] D. M. Bylander and L. Kleinman, Phys. Rev. B **41**, 7868 (1990).
[13] C. B. Geller, W. Wolf, S. Picozzi, A. Continenza, R. Asahi, W. Mannstadt, A. J. Freeman, and E. Wimmer, Appl. Phys. Lett. **79**, 368 (2001).
[14] J. Heyd, J. E. Peralta, G. E. Scuseria, and R. L. Martin, J. Chem. Phys. **123** (2005).
[15] I. C. Gerber, J. G. Angyan, M. Marsman, and G. Kresse, J. Chem. Phys. **127**, 054101 (2007).
[16] A. Savin, in *Recent Advances in Density Functional Methods Part I*, edited by D. P. Chong (World Scientific, Singapore, 1995), p. 129.
[17] T. Leininger, H. Stoll, H.-J. Werner, and A. Savin, Chem. Phys. Lett. **275**, 151 (1997).
[18] H. Iikura, T. Tsuneda, T. Yanai, and K. Hirao, J. Chem. Phys. **115**, 3540 (2001).
[19] R. Baer and D. Neuhauser, Phys. Rev. Lett. **94**, 043002 (2005).
[20] E. Livshits and R. Baer, Phys. Chem. Chem. Phys. **9**, 2932 (2007).
[21] E. Livshits and R. Baer, J. Phys. Chem. A **112**, 12789 (2008).
[22] T. Gershon, R. Baer, and L. Kronik, work in progress (2008).
[23] J. E. Robinson, F. Bassani, R. S. Knox, and J. R. Schreiffer, Phys. Rev. Lett. **9**, 215 (1962).
[24] T. Yanai, D. P. Tew, and N. C. Handy, Chem. Phys. Lett. **393**, 51 (2004).
[25] T. Stein, L. Kronik, and R. Baer, J. Am. Chem. Soc. **in press** (2009).
[26] S. Baroni, A. Dal Corso, d. G. S., P. Giannozzi, C. Cavazzoni, G. Ballabio, S. Scandolo, G. Chiarotti, P. Focher, A. Pasquarello, K. Laasonen, A. Trave, R. Car, N. Marzari, and KokaljA., http://www.pwscf.org/.
[27] J. P. Perdew and A. Zunger, Phys. Rev. B **23**, 5048 (1981).
[28] K. Chakrabarti, V. K. Mathur, and R. J. Abbundi, Phys. Rev. B **39**, 10406 (1989).
[29] V. I. Siklitsky, New Semiconductor Materials. Characteristics and Properties **http://www.ioffe.rssi.ru/SVA/NSM** (2005).
[30] A. DalCorso, F. Mauri, and A. Rubio, Phys. Rev. B **53**, 15638 (1996).
[31] Y. Goldberg, edited by M. E. Levinshtein, S. L. Rumyantsev and M. S. Shur (John Wiley & Sons Inc., New York, 2001), p. 31.
[32] I. Akasaki and Hashimot.M, Sol. Stat. Comm. **5**, 851 (1967).
[33] B. W. Henvis and M. Haas, Journal of Physics and Chemistry of Solids **23**, 1099 (1962).
[34] P. A. Thiry, M. Liehr, J. J. Pireaux, and R. Caudano, Phys. Rev. B **29**, 4824 (1984).
[35] C. Kittel, *Introduction to solid state physics* (Wiley, New York,, 1966).
[36] I. Vurgaftman, J. R. Meyer, and L. R. Ram-Mohan, J. Appl. Phys. **89**, 5815 (2001).
[37] Vanvecht.Ja, Phys. Rev **182**, 891 (1969).
[38] American Institute of Physics. and D. E. Gray, *American Institute of Physics handbook. Section editors: Bruce H. Billings [and others] Coordinating editor: Dwight E. Gray* (McGraw-Hill, New York,, 1972).
[39] T. Soma, Journal of Physics C-Solid State Physics **11**, 2669 (1978).
[40] A. Polian and M. Grimsditch, Phys. Rev. B **60**, 1468 (1999).
[41] E. S. Zouboulis and M. Grimsditch, Journal of Geophysical Research-Solid Earth and Planets **96**, 4167 (1991).
[42] S. Speziale, C. S. Zha, T. S. Duffy, R. J. Hemley, and H. K. Mao, Journal of Geophysical Research-Solid Earth **106**, 515 (2001).
[43] A. Dewaele, G. Fiquet, D. Andrault, and D. Hausermann, Journal of Geophysical Research-Solid Earth **105**, 2869 (2000).
[44] B. P. Singh, S. K. Srivastava, and K. Dinesh, Physica B: Condensed Matter **349**, 401 (2004).
[45] H. Spetzler, C. G. Sammis, and R. J. Oconnell, Journal of Physics and Chemistry of Solids **33**, 1727 (1972).
[46] A. Rubio, J. L. Corkhill, M. L. Cohen, E. L. Shirley, and S. G. Louie, Phys. Rev. B **48**, 11810 (1993).
[47] M. S. Hybertsen and S. G. Louie, Phys. Rev. B **34**, 5390 (1986).
[48] M. S. Hybertsen and S. G. Louie, Phys. Rev. Lett. **55**, 1418 (1985).
[49] M. Rohlfing, P. Krüger, and J. Pollmann, Phys. Rev. B **48**, 17791 (1993).
[50] R. W. Godby, M. Schlueter, and L. J. Sham, Phys. Rev. Lett. **56**, 2415 (1986).
[51] R. W. Godby, M. Schlueter, and L. J. Sham, Phys. Rev. B **37**, 10159 (1988).
[52] F. Gygi and A. Baldereschi, Phys. Rev. Lett. **62**, 2160 (1989).
[53] X. Zhu and S. G. Louie, Phys. Rev. B **43**, 14142 (1991).
[54] M. Usuda, N. Hamada, K. Shiraishi, and A. Oshiyama, Physica Status Solidi C **0**, 2733 (2003).
[55] I. N. Remediakis and E. Kaxiras, Phys. Rev. B **59**, 5536 (1999).
[56] L. Fritsche and Y. M. Gu, Phys. Rev. B **48**, 4250 (1993).
[57] M. Städele, M. Moukara, J. A. Majewski, P. Vogl, and A. Görling, Phys. Rev. B **59**, 10031 (1999).